\documentstyle[11pt,newpasp,twoside]{article}
\def\HI{H{\small I}}
\def\slHI{\sl H\,\small\sl I}
\def\oii{[O{\small II}]}
\def\oiii{[O{\small III}]}
\def\ltae{\raisebox{-0.6ex}{$\,\stackrel
{\raisebox{-.2ex}{$\textstyle <$}}{\sim}\,$}}

\def\edcomment#1{\iffalse\marginpar{\raggedright\sl#1\/}\else\relax\fi}
\reversemarginpar
\input{psfig}
\begin{document}
\title{HI absorption in  radio galaxies}
\author{R. Morganti, T.A. Oosterloo}
\affil{Netherlands Foundation for Research in Astronomy, Postbus 2, 7990 AA,
Dwingeloo, NL}
\author{G. van Moorsel}
\affil{National Radio Astronomy Observatory, Socorro,
NM 87801, USA}
\author{C.N. Tadhunter}
\affil{Dept. Physics, University of Sheffield, Sheffield S3 7RH, UK}
\author{N. Killeen}
\affil{Australia Telescope National Facility, CSIRO, P.O. 
  Box 76, 2121 Epping NSW, Australia}

\begin{abstract}
Twenty-two powerful radio galaxies have been searched for \HI\ absorption.  We
find the highest probability of detecting \HI\ in absorption  among
narrow-line compact (or small) galaxies or galaxies with indication of richer
interstellar medium (i.e.  with ongoing or recent star-formation).  We discuss
the difficulty in the interpretation of the origin of the \HI\ absorption due to
the uncertainty in the systemic velocity of the galaxies.  \end{abstract}

%\section{Introduction}

A number of interesting features and phenomena are taking place in the central
regions of Active Galactic Nuclei (AGN).  Among others, and relevant for this
work, are: i) the presence of nuclear tori (on the pc-scale) and larger scale
disk/tori (extending up to kpc scale and beyond), ii) the possible presence of
in-falling gas that has been often suggested as responsible for
``feeding'' of the central engine, iii) the presence of gas disturbed by the
interaction with the radio plasma and finally, iv) the presence of
larger-scale tails of gas (that in particular conditions can be seen projected
against the nuclear regions) possibly the result of merger or interaction with
an other object.

The study of all these phenomena requires kinematical information about the
gas and therefore one of the few ways to obtain this is by studying the \HI\
absorption against the nuclei of radio galaxies.  The advantage of this
technique is that observations of the \HI\ absorption against radio continuum
sources allow to detect relatively small quantities of \HI\ even in high
redshift objects.

\section{A well-studied sample}

Many of the cases of \HI\ absorption detected so far in radio galaxies have
been interpreted as evidence for obscuring tori.  Thus it can be interesting
to investigate whether the presence of absorption correlates with other
orientation-dependent indicators (like, e.g., the width of the optical lines). 
In order to do this, we have been looking for \HI\
absorption in a sample of radio galaxies for which a wealth of radio, optical
and X-ray information is available. 

The observed radio galaxies have been selected from the 2-Jy sample of radio
sources: a summary of the data available on this sample can be found in
Tadhunter et al.  (1998) and Morganti et al. (1999) and references therein.  This
sample includes strong (S$_{\rm 2.7GHz}>$ 2 Jy) radio sources with declination
$\delta<10^\circ$.  So far we have look for \HI\ absorption
in 22 radio galaxies.  The observed objects have been selected to have a
strong enough radio core and $z \ltae 0.2$, i.e.  the highest redshift
reachable by most currently available 21-cm radio receivers, although
the sample is complete up to $z = 0.7$.  Due to the southern declination of
most objects in the sample, only an handful of higher redshift objects
could be observed with the UHF system in use at the WSRT.  The role of GMRT in
extending this study at higher redshift will be essential.  The observations
so far were done mainly with the Australia Telescope Compact Array (ATCA) and
with the Very Large Array (VLA) (depending on the declination of the objects). 

%In this paper we have set H$_\circ$ = 50 and q$_\circ$=0.

\section{Occurrence  and  origin of the HI absorption}

Of the 22 radio galaxies for which the data have been analysed so far,
we find five galaxies with clear \HI\ absorption (with optical depths
ranging between 2 and 10 \%): PKS~1318--43 (NGC~5090), PKS~1549--79,
PKS~1717--00 (3C353), PKS~1814--63 and PKS~2314+03 (3C459).  For the
undetected galaxies, a 3$\sigma$ limit to the optical depth of at most
few percent has been obtained in most cases. 
Here are more details about the occurrence of the \HI\ absorption.

-- Of the {\sl 11 Fanaroff-Riley (FR) type-I observed, we detected only one}
(PKS~1318-43).  A few more can be included if the data available in literature
are considered (e.g.  Hydra~A, NGC~4261 etc.).  

-- So far, in our sample {\sl no \slHI\ absorption has been detected in
broad line radio galaxies (BLRG)} (see also below). 

-- {\sl The three compact (or small) radio galaxies observed by us were all
detected}: PKS~1814-63, PKS~1549-79, 3C459.  The \HI\ absorption profiles of
the first two are shown in Fig.1.  Two more compact sources in the sample were
observed with the WSRT (R.  Vermeulen private communication).  Only one of the
two has been detected (PKS~0023--26).  However, it is worth mentioning that
the undetected source (PKS~2135--20) is a broad line radio galaxy, consistent
with the trend for BLRG described above.

-- Only two radio galaxies (3C459 and PKS~1549-79) in the whole 2-Jy sample
show evidence of a star-burst or young stellar population component in their
optical spectra (in addition to the old stellar population and the non-thermal
power law component).  These galaxies, seems to have a richer
interstellar medium (ISM), they are also bright in FIR emission, compared to
the other powerful radio galaxies in the sample and  {\sl \slHI\ absorption has been
detected in both of them}.

\medskip

The low detection rate of \HI\ absorption in FR type-I galaxies is somewhat
surprising given that, as shown from recent HST images, gas and dust, in
particular circumnuclear dust-lanes, are commonly present in low-luminosity
radio galaxies (see e.g. Chiaberge et al.  1999).  However, also
from optical studies it is clear that the  nuclei of FR type-I galaxies 
appear basically not
obscured.  Indeed, for this type of radio galaxies the presence of thick 
disks surrounding the central black hole is not really required by
unified schemes (Urry \& Padovani 1995).  Thus, our result is consistent with
the work of Chiaberge et al.  (1999) arguing that the ``standard'' pc scale,
geometrically thick torus is not present in low-luminosity radio galaxies.  A
possibility is that the circumnuclear disks in these objects are typically
thin ($\ltae 20$ pc) as found in the case of NGC~4261 (Jaffe \& McNamara
1994).

% Fig. 1
\begin{figure}
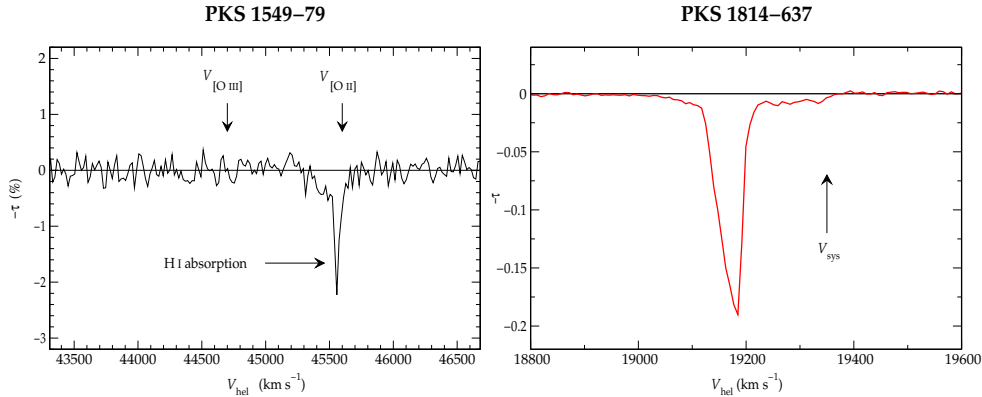

\centerline{\psfig{figure=pks1549-prof.eps,height=5.2cm,angle=-90}
\psfig{figure=pks1814Prof.eps,height=5.2cm,angle=-90}}
\caption{{\it Left:} HI absorption spectra (from the ATCA) observed in the
source PKS 1549-79.  Two different velocities have been derived from the
\oiii\ and \oii\ emission lines.  See text for details.  {\it Right:} HI
absorption spectra (from the ATCA) observed in the source PKS~1814-63.  The
new systemic velocity is indicated.}
\end{figure}

In the case of FR type-II radio galaxies, the presence of a thick disk
is a vital ingredient in the unified schemes hypothesis.  Therefore,
\HI\ absorption is expected in narrow line galaxies.  It appears
that for powerful radio galaxies the highest probability of detecting
\HI\ in absorption is among narrow-line {\sl compact} (or small)
galaxies (as already pointed out by Conway 1996) or galaxies with a
richer interstellar medium (i.e.  with ongoing or recent
star-formation).  Unfortunately, because of the selection criteria, our
sample is biased against powerful, {\it extended}, narrow line
radiogalaxies.  This bias prevent us to establish whether the \HI\
absorption is common also in the extended narrow line radio galaxies or
only in the compact ones. 

So far, \HI\ absorption seems absent in BLRG.  If the \HI\ absorption is
due to an obscuring torus or disk, this result is what na\"\i vely is
expected according to the unified schemes.  Broad line radio galaxies
are in fact supposed to be galaxies seen pole-on and therefore
obscuration from the torus should not occur. 

Is the absorbing material really situated in a disk/torus? From the average profile
we cannot say for sure but the presence of the absorption in narrow-line
galaxies and the absence in broad line galaxies seems to be consistent with
this possibility.  On the other hand the presence of absorption in the case of
richer ISM could make the story more complicated. 

In order to really answer this question, {\sl imaging of the \HI\ absorption
on the VLBI scale} together with {\sl good values for the systemic velocities}
are needed.

\section{The problem of the systemic velocity}
 
It was already pointed out by Mirabel (1989) how the systemic velocities
derived from emission lines can be both uncertain and biased by motions
connected with the gas producing these emission lines. 

We have found a very good example of this uncertainty in one of the objects
showing \HI\ absorption: PKS~1549-79.  In this galaxy the available redshift
derived from \oiii\ emission lines (that is commonly used for the majority of radio galaxies)
would suggest the \HI\ being strongly redshifted (see Fig.1).  A more detailed
analysis of the spectrum has shown that the \oii\ lines give a velocity about
800 km/s higher compared to what derived from the \oiii\ lines and {\sl
consistent with the \slHI\ absorption} (see Fig.~1).  What should we consider as
systemic velocity of the galaxy? The \oiii\ lines appear to be very broad, one
of the broadest observed in radio galaxies with a FWHM of $\sim 1300$
km/s. Thus, 
we belive that much of the \oiii\ is emitted by an inner narrow line
region, which is undergoing {\sl outflow} from the nucleus, whereas the
\oii\ are  likely to be closer to the real systemic
velocity of this galaxy.

Also in the case of the other compact object PKS~1814-637 a more accurate
redshift has been derived.  The \HI\ absorption appears now to be mainly
blue-shifted (unlike from the value derived using the old systemic velocity)
as shown in Fig.  1.  With this new piece of information the \HI\ absorption,
and in particular the broad component visible at about 1\% level optical
depth, could be due to some gas outflow connected perhaps to interaction
between the radio plasma and the ISM.  \HI\ absorption due to gas outflow has
been clearly seen already in few objects as the Seyfert galaxy IC~5063
(Oosterloo et al.  1999), the superluminal source 3C216 (Pihlstr\"om et
al.1999) and 3C326 (Conway et al. in prep). 
 
So far, we have not found objects with clear redshifted \HI\ absorption.


\begin{references}

\reference{Chiaberge M., Capetti A., Celotti A., 1999, A\&A 349, 77}
\reference{Conway J.E. 1996, in {\it The second Workshop on GPS and CSS
 Radio Sources}, Eds. Snellen, Schilizzi et
al. M.N. Publ JIVE, Leiden p.198}
%\reference{de Koff et al. }
%\reference{Dickson R. 1997, Ph.D. Thesis Sheffield University}
%\reference{Dwarakanath K.~S., Owen F.~N.  \& van Gorkom J.~H.  1995, \apj, 442,
%L1}
\reference{Jaffe W. \& McNamara B. 1994, ApJ, 434, 110}
\reference{Mirabel I.F. 1989, \apj\ 340, L13}
\reference{Morganti R., Oosterloo T.A., Tadhunter C.N., Aiudi R.,
Jones P., Villar-Martin 1999, A\&A Suppl. 140, 355}
\reference{Oosterloo T.A., Morganti R., Tzioumis A., Reynolds J., King E.,
McCulloch P., Tsvetanov Z. 1999, \aj\ in press}
\reference{Pihlstr\"om Y.M., Vermeulen R.C., Taylor G.B., Conway J.E. 1999,
\apj\ 525, L13}
\reference{Tadhunter C.N., Morganti R., Robinson A., Dickson R., Villar-Martin
M.  \& Fosbury R.A.E., 1998, \mnras\ 298, 1035}
%\reference{van Langevelde H.J., Pihlstr\"om Y.M., Conway J.E., Jaffe W.,
%Schilizzi R.T. 1999, A\&A in press}

\end{references}
\end{document}